\documentclass[12pt]{article}
\usepackage{graphicx}
\usepackage{cite}
\usepackage{bm}
\usepackage{amssymb}
\usepackage{amsmath}
\usepackage{amsthm}
\usepackage{fullpage}
\DeclareMathOperator\erfc{erfc}

\begin{document}

\title{\textbf{Low-Complexity Concatenated}\\\textbf{LDPC-Staircase Codes}}

\markboth{IEEE/OSA J. Lightwave Technology}{Barakatain and Kschischang}

\author{Masoud Barakatain and Frank R. Kschischang\thanks{The authors are with the
Edward S. Rogers Sr.\ Dept.\ of Electrical and Computer Engineering,
University of Toronto, Toronto, ON M5S 3G4, Canada.  Email:
\texttt{\{barakatain,frank\}@ece.utoronto.ca}.  To appear in
\emph{J. Lightwave Technology}, 2018.}
}

\date{September 21, 2018}

\maketitle

\begin{abstract}
A low-complexity soft-decision concatenated FEC scheme, consisting of an
inner LDPC code and an outer staircase code is proposed. The inner code is
tasked with reducing  the bit error probability below the outer-code
threshold. The concatenated code is obtained by optimizing the degree
distribution of the inner-code ensemble to minimize estimated data-flow,
for various choices of outer staircase codes. A key feature that emerges
from this optimization is that it pays to leave some inner codeword bits
completely uncoded, thereby greatly reducing a significant portion of the
decoding complexity. The trade-off between required SNR and decoding
complexity of the designed codes is characterized by a Pareto frontier.
Computer simulations of the resulting codes reveals that the net
coding-gains of existing designs can be achieved with up to 71\% reduction
in complexity. A hardware-friendly quasi-cyclic construction is given for
the inner codes, which can realize an energy-efficient decoder
implementation, and even further complexity reductions via a layered
message-passing decoder schedule.
\end{abstract}

\section{Introduction}
\label{sec:introduction}

Recent
optical transport network (OTN) system proposals increasingly
specify the use of  soft-decision codes, i.e., codes that can make use of
probabilistic symbol reliabilities, in the forward error correction (FEC)
scheme. At a similar overhead (OH) and signal-to-noise ratio (SNR),
soft-decision codes can achieve coding gains of 1 dB, or more, relative to
the hard-decision codes used in earlier OTN proposals
\cite{Tzimpragos2016}. The excellent performance of soft-decision codes
comes, however, at the expense of a  significantly increased decoding
complexity. A comparison of the implementations of soft- and hard-decision
decoders shows that soft-decision decoders typically consume an order of
magnitude more power than hard-decision decoders
\cite{Weiner2014,Ou2014,Youngjoo2013,Hoyoung2014} operating at the same OH
and throughput. 

This paper continues the work of Zhang and Kschischang
\cite{ZhangKschischang2017} on designing low-complexity concatenated FEC
schemes for applications with high throughput. Their design consists of an
inner soft-decision low-density generator-matrix (LDGM) code concatenated
with an outer hard-decision staircase code. The degree distribution of the
inner LDGM code ensemble is obtained by solving an optimization problem,
minimizing the estimated data-flow of the inner-code decoder, while
searching a table of staircase codes to find the optimal inner and outer
code pair. At 20\% OH, the codes proposed in \cite{ZhangKschischang2017}
can achieve up to 46\% reduction in complexity, when compared with other
low-complexity designs.

In this paper, we adopt the concatenated FEC structure of
\cite{ZhangKschischang2017}, but we consider a different ensemble of inner
codes. The task of the inner code, similar to that of
\cite{ZhangKschischang2017}, is to reduce the bit error rate (BER) of the
bits transferred to the outer staircase code to below the threshold, which
enables the outer code to take the BER further down, below  $10^{-15}$, as
required by OTNs.  We re-design the inner code to further reduce its
data-flow, thereby achieving an FEC solution with even lower complexity
than the codes reported in \cite{ZhangKschischang2017}. 

A key characteristic that emerges from the re-designed inner-code
optimization is that some inner codeword bits remain uncoded! These bits
bypass the inner code, and are protected only by the outer-code. We propose
a method to analyze and optimize the inner-code ensemble, and show that the
resulting codes can reduce the inner-code data-flow by up to 71\%, when
compared to \cite{ZhangKschischang2017}. We show that, when block length is
sufficiently large, codes generated according to the obtained inner-code
ensembles perform as expected, verifying the design approach. 

To realize a pragmatic decoder implementation, we construct quasi-cyclic
(QC) codes of practical length, generated according to the obtained
inner-code ensembles. We show that the performance of randomly-generated
inner codes of large block-length can be achieved by QC codes of practical
length on the order of 6000 to 15000. A QC-structured inner code allows for
decoder hardware implementations that are very energy efficient
\cite[Ch.~3]{Milicevic2017Low}. The QC structure also enables a layered
message-passing decoding schedule. We show that, compared with the flooding
schedule, layered decoding of the QC-structured codes reduces the
complexity by up to 50\%.

The rest of this paper is organized as follows. In
Sec.~\ref{sec:innercodestructure} we describe the inner-code structure,
code parameters, and complexity measure. In
Sec.~\ref{sec:complexityoptimizeddesign} we describe how extrinsic
information transfer (EXIT) functions can be used to predict the inner-code
performance, and we describe the inner-code optimization procedure. In
Sec.~\ref{sec:results} we present simulation results and give a
characterization of the trade-off between the required SNR and decoding
complexity for the concatenated code designs. Designs with QC-structured
codes are also discussed in Sec.~\ref{sec:results}, and a comparison  with
existing soft-decision FEC solutions is presented. In
Sec.~\ref{sec:conclusion} we provide concluding remarks.

Throughout this paper, we consider signaling using a Gray-labeled
quadrature phase-shift keying constellation, with unit energy per symbol.
We assume a memoryless, additive white Gaussian noise (AWGN) channel, with
covariance matrix $\sigma^2\bm{I}$, where $\bm{I}$ is the $2 \times 2$
identity matrix. In this setting, SNR, in decibels, is denoted by $E_s/N_0
\triangleq -10 \log_{10}{2\sigma^2}$.

\section{The Inner-Code Structure}
\label{sec:innercodestructure}

\subsection{Code Description}
\label{subsec:codedescription}

We use low-density parity-check (LDPC) codes as inner codes. A significant
feature of the inner-code ensemble is that we allow for both degree-zero
and degree-one variable nodes. Degree-zero variable nodes are uncoded, and
thus incur zero inner decoding complexity. Also, as will be discussed in
Sec.~\ref{subsec:complexitymeasure}, degree-one variable nodes do not add
to the data-flow throughout the decoding procedure, thus they also incur no
inner decoding complexity.

\begin{figure}
\centering
\includegraphics{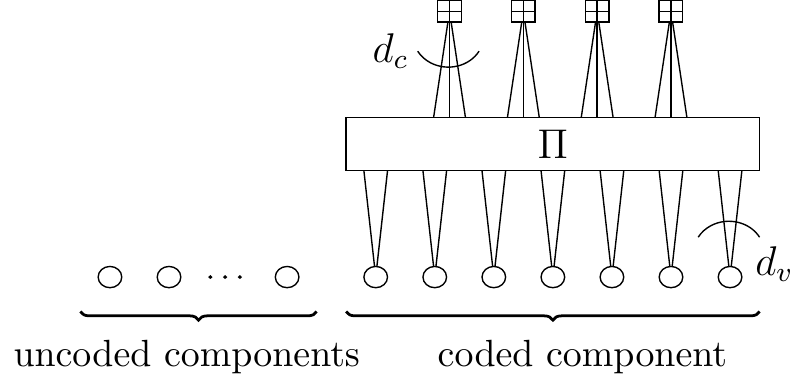}
\caption{Tanner graph of an inner code, consisting of some degree-zero
variable nodes (uncoded components) and a coded component. The rectangle
labeled by $\Pi$ represents an edge permutation.}
\label{fig:tanner}
\end{figure}

We denote the inner code by $\mathcal{C}_{\text{in}}$ and its rate by
$R_{\text{in}}$.  The coded component (excluding the uncoded bits) form an
LDPC code of length $n_c$ and rate $R_c$.  A Tanner graph for a member of
the inner-code ensemble is sketched in Fig.~\ref{fig:tanner}. 

Note that the LDGM inner code of \cite{ZhangKschischang2017} is an instance
of the ensemble defined above. However, in an LDGM code, check nodes are
associated randomly with variable nodes, inducing a Poisson distribution on
variable-node degrees. In this work, the variable-node degree distribution
is carefully optimized to achieve small decoding complexity.

\subsection{Ensemble Parameterization}
\label{subsec:ensembleparameterization}

The inner code ensemble is described by its variable-node and check-node
degree distributions. We denote the maximum degree of a variable node or a
check node by $D_v$ and $D_c$, respectively. We define the normalized,
node-perspective, degree distributions as $L(x) \triangleq \sum_{i=0}^{D_v}
L_i x^i \text{, } R(x) \triangleq \sum_{i=2}^{D_c} R_i x^i$, where $L_i$ is
the fraction of variable nodes that have degree $i$, and $R_i$ is the
fraction of check nodes that have degree $i$.

We define the normalized, edge-perspective, degree distributions as
$\lambda(x) \triangleq L'(x)/L'(1) = \sum_{i=1}^{D_v} \lambda_i x^{i-1} $
and $ \rho(x) \triangleq R'(x)/R'(1)=\sum_{i=1}^{D_c} \rho_i x^{i-1}$,
corresponding to the variable and check nodes, respectively, where
$L'(x)=\frac{d}{dx}L(x)$ and $R'(x)=\frac{d}{dx}R(x)$.

The portion of uncoded bits is given by $L_0$, thus the coded component
rate, $R_c$, satisfies
\begin{equation}
R_{\text{in}} = L_0 + R_c(1-L_0). 
\label{eqn:innr rate1}
\end{equation}
For computational simplicity, we only consider check degree
distributions concentrated on one or two consecutive degrees.
For a check-node degree distribution that is
concentrated on a single degree, $d_c$, $R_c$ is related to the
edge-perspective variable degree distribution by
\begin{equation}
\sum_{i=1} ^ {D_v}\frac{\lambda_i}{i} = \frac{1}{d_c(1-R_c)}.
\label{eqn:innr rate2}
\end{equation}
Therefore, for a given inner-code rate $R_{\text{in}}$, the inner-code
ensemble is completely described by the pair $( L_0 ,  \Lambda)$, where
$\Lambda = ( \lambda_1,  \lambda_2, \ldots,  \lambda_{D_v} )$. We refer to
the pair $( L_0 ,  \Lambda)$ as the \textit{design parameters}.

For reasons described in Sec.~\ref{subsec:complexitymeasure}
and Sec.~\ref{subsec:exitchartanalysis},
degree-one
variable nodes receive special treatment in our design. We define $\nu$ to
be the average number of degree-one variable nodes connected to each check
node. In terms of the code parameters, $\nu$ can be expressed as
\begin{equation}
\label{eqn:nu}
\nu = d_c \lambda_1.
\end{equation}

For a check-degree distribution that is concentrated on two consecutive
degrees, $d_c$ and $d_c+1$, the edge-perspective check-degree distribution
$\rho(x)$ is specified by fixing the average check-node degree $\bar{d_c}$,
and is obtained as
\begin{equation*}
\rho(x) = \frac{d_c(d_c+1-\bar{d_c})}{\bar{d_c}}x^{d_c-1} + \frac{\bar{d_c} - d_c(d_c+1-\bar{d_c})}{\bar{d_c}}x^{d_c},
\end{equation*}
where $d_c = \lfloor \bar{d_c} \rfloor$. In this case, equation (2) and all
other following equations still hold, when $d_c$ is replaced with
$\bar{d_c}$.

\subsection{Complexity Measure}
\label{subsec:complexitymeasure}

We use the same complexity measure described in
\cite[Eq.~4]{ZhangKschischang2017}, to quantify, and eventually minimize, the
required data-flow at the decoder. The concatenated code decoder complexity is
defined as 
\begin{equation}\label{eqn:eta}
\eta = \frac{ \eta_i }{R_{\text{sc}}} + P,
\end{equation}
where $\eta_i$ is the inner code complexity score, $R_{\text{sc}}$ is the outer
staircase code rate, and $P$ is the number of post-processing operations per
information bit at the outer-code decoder.
In this paper we have set $P=0$, since the decoding complexity, per bit,
of the staircase code is typically two to three orders of magnitude smaller
than that of the inner code.  This can be estimated as follows for
the rate 15/16 staircase code with a (1408,1364) component code.  Typically,
each component code is ``visited'' by the iterative decoder about
four times during the decoding (where the decoding, i.e., processing
of a syndrome, is performed using a small table-lookup-based circuit).
Since each information bit is protected by two component codes, the
average number of bits recovered per decoding attempt is 170.5, giving
a complexity of $P \approx 0.006$ decoding attempts per decoded bit.

The complexity score of the inner-code, $\eta_i$, can be computed as
\begin{equation}\label{eqn:icomp}
\eta_i = \frac{ (1-R_{\text{in}}) (d_c - \nu) I}{R_{\text{in}}},
\end{equation}
where $I$ is the maximum number of decoding iterations allowed for the inner-code decoder. Note that, similar to \cite{ZhangKschischang2017}, the complexity score in (\ref{eqn:icomp}) does not account for messages of degree-one variable nodes, as they remain constant throughout the decoding procedure.

\section{Complexity-optimized Design}
\label{sec:complexityoptimizeddesign}

\subsection{EXIT chart analysis}
\label{subsec:exitchartanalysis}

We analyze the coded component of the inner code using a version of EXIT
functions
\cite{ArdakaniKschischang2004,SmithArdakani2010}.
Under the assumption that the all-zero codeword is transmitted, we define
the error-probability EXIT function $f_{\Lambda}$, that takes
$p^{\text{in}}$, the probability of error in messages coming from the
variable nodes, as input, and outputs $p^{\text{out}}$, the probability of
error in messages coming from the variable nodes, after one round of
sum-product message-passing, i.e.,
\begin{equation}\label{eqn:pout}
p^{\text{out}} = f_{\Lambda}(p^{\text{in}}).
\end{equation}

Using the law of total probability, we can write $p^{\text{out}}$ as
\begin{equation}\label{eqn:totprob}
p^{\text{out}} = \sum_{i=1}^{D_v} \lambda_i p_i^{\text{out}},
\end{equation}
where $p_i^{\text{out}}$ is the probability of error in messages coming
from a degree-$i$ variable node. From (\ref{eqn:pout}) and
(\ref{eqn:totprob}) we get
\begin{equation}\label{eqn:poutfinal}
p^{\text{out}} = f_{\Lambda}(p^{\text{in}}) = \sum_{i=1}^{D_v} \lambda_i f_{i,\Lambda}(p^{\text{in}}),
\end{equation}
where functions $f_{i,\Lambda}$ are called \emph{elementary} EXIT
functions. Function $f_{i,\Lambda}$ takes $p^{\text{in}}$ as an argument,
and produces $p^{\text{out}}_i$, the probability of error in messages
coming from the degree-$i$ variable nodes, after one round of sum-product
message-passing. As shown in \cite{ArdakaniKschischang2004}, in practice the
elementary EXIT charts' dependence on $\Lambda$ can be neglected.
Therefore, (\ref{eqn:poutfinal}) can be written as
\begin{equation}
p^{\text{out}} = f(p^{\text{in}}) = \sum_{i=1}^{D_v} \lambda_i f_i(p^{\text{in}}).
\end{equation}

In \cite{ArdakaniKschischang2004} a method is proposed, that, for an LDPC
code ensemble without degree-zero and degree-one variable nodes,
approximates the elementary EXIT charts using Monte-Carlo simulation.
Assuming that the messages coming from the variable nodes have a symmetric
Gaussian distribution with mean $m= (2\erfc^{-1}(p^{\text{in}}))^2$ and
variance $\sigma^2 = 2m$, an empirical distribution for check-node messages
is generated by performing the check-node computation on samples of
variable-node messages. A degree$-i$ variable node then adds its channel
message and $i-1$ independent samples of check-node messages, to generate
a sample of $f_i(p^{\text{in}})$. It is shown that the elementary EXIT
charts generated by interpolating the average of a large number
of $f_i(p^{\text{in}})$ samples closely replicate the actual
elementary EXIT charts.

In our design, however, we must take into account the presence of
degree-one variable nodes in obtaining the elementary EXIT charts with the
method of \cite{ArdakaniKschischang2004}, as the messages from such nodes
significantly affect the check-node operation.  To this end, we generate
the elementary EXIT charts for a pre-set value of $\nu$, the average number
of degree-one variable nodes connected to each check node, as defined in
(\ref{eqn:nu}).  In the Monte-Carlo simulation described above, we modify
the check-node operation to account for the fact that each check node is
connected to, on average, $\nu$ degree-one variable nodes, and therefore
receives only their channel observation.  In particular, given a value of
$\nu$, let $\theta \in [0,1)$ satisfy $\theta \lfloor \nu \rfloor +
(1-\theta) \lceil \nu \rceil = \nu$.  We then assume that a fraction
$\theta$ of the check nodes are connected to $\lfloor \nu \rfloor$
degree-one variable nodes and the remainder are connected to $\lceil \nu
\rceil$ degree-one variable nodes.

Note that the SNR,  $d_c$, and $\nu$  are the only parameters needed to
compute the elementary EXIT charts. Since they do not depend on inner-code
design parameters, elementary EXIT charts can be pre-computed. Therefore,
when SNR, $d_c$, and $\nu$ are given, the problem of  inner-code design
reduces to the problem of appropriately shaping an EXIT chart out of its
constituent elementary EXIT charts. 

\subsection{Code Optimization}
\label{subsec:codeoptimization}

Similar to \cite{ZhangKschischang2017}, we view the problem of designing
the concatenated FEC scheme as a multi-objective optimization with the
objectives $(E_s/N_0,\eta_i)$. In both parameters, smaller is better, i.e.,
we wish to minimize the SNR needed to achieve the target error rate and we
wish to minimize the estimated complexity needed to do so. Given a
concatenated code rate, $R_{\text{cat}}$, we characterize the trade-off
between the objectives by finding their Pareto frontier. For any SNR, we
find a pair (if it exists), consisting of an outer staircase code and an
inner-code ensemble, with minimum complexity, that together, bring the BER
below $10^{-15}$.

Concatenated code optimization procedure is as follows. When an
$R_{\text{cat}}$ is specified,  we loop  over a table of staircase codes
such as \cite[Table 1]{ZhangKschischang2017}. Recall that each staircase
code specifies $R_{\text{sc}}$ and $p_{\text{sc}}$, the rate and threshold
of the outer code, respectively.  For each staircase code, we perform the
inner-code ensemble complexity optimization.

It is shown in \cite{SmithArdakani2010}, that, given the EXIT function, the
number of iterations, $I$, required by the inner-code coded component, to
take the  variable nodes \textit{message} error probability from $p_0$, the
channel BER, down to $p_t$, a target \textit{message} error probability,
can be closely approximated as
\begin{equation}\label{eqn:itapprox}
I \approx \int_{p_t}^{p_0} \frac {\text{d}p}{p \log\left( \frac{p}{f_{\Lambda}(p)} \right)}.
\end{equation}

The target \textit{information} bit error probability for the coded
component, $P_t$, can be computed from $p_t$ and $\Lambda$ as described in
\cite{SmithArdakani2010}, and should satisfy
\begin{equation}\label{eqn:L0}
L_0p_0 + (1-L_0)R_cP_t \leq p_{\text{sc}}R_{\text{in}}.
\end{equation}
Note that the maximum feasible $L_0$, which we denote by
$L_0^{\text{max}}$, can be obtained from (\ref{eqn:L0}), by setting
$P_t=0$, as
\begin{equation}\label{eqn:L0max}
L_0^{\text{max}} = \frac{ p_{\text{sc}}R_{\text{in}}}{p_0}.
\end{equation}

From (\ref{eqn:icomp}) and (\ref{eqn:itapprox}), the complexity-optimized
inner-code ensemble is obtained by solving the following optimization
problem:
\begin{align}
\underset{(L_0,\Lambda)}{\text{minimize}} \ \ & \ \frac{(1-R_{\text{in}})(d_c - \nu)}{R_{\text{sc}}R_{\text{in}}} \int_{p_t}^{p_0} \frac {\text{d}p}{p \log\left( \frac{p}{f_{\Lambda}(p)} \right)}, \label{eqn:opt} \\
\text{subject to} \ &\sum_{i=1} ^ {D_v}\frac{\lambda_i}{i} \geq \frac{1-L_0}{d_c(1-R_{\text{in}})},\label{eqn:rateconstraint}\\
&\sum_{i=1} ^ {D_v} \lambda_i = 1,\label{eqn:unitconstraint}\\
&\lambda_1 d_c = \nu,\label{eqn:nuconstraint}\\
&0 \leq \lambda_i \quad \forall i \in \{1, \dots, D_v\},\label{eqn:posconstraint}\\
&0 \leq L_0 \leq L_0^{\text{max}},\label{eqn:L0constraint}\\
&f_{\Lambda}(p) < p  \quad \forall p \in [p_t,p_0] ,\label{eqn:openconstraint}\\
&L_0p_0 + (1-L_0)R_cP_t \leq p_{\text{sc}}R_{\text{in}}.\label{eqn:targetconstraint}
\end{align}

In this optimization problem formulation, constraint
(\ref{eqn:rateconstraint}) ensures that the obtained complexity-optimized
code has the desired rate (see (\ref{eqn:innr rate1}) and (\ref{eqn:innr
rate2})). Constraints (\ref{eqn:unitconstraint})--(\ref{eqn:L0constraint})
ensure the validity of the obtained ensemble. Constraint
(\ref{eqn:openconstraint}) ensures that the obtained EXIT-curve remains
\textit{open} throughout the decoding procedure, i.e., for all $p \in
[p_t,p_0]$. Finally, (\ref{eqn:targetconstraint}) ensures that the
inner-code output BER is at or below the outer-code threshold.

Note that, in terms of the optimization parameters, constraints
(\ref{eqn:rateconstraint})--(\ref{eqn:openconstraint}) are linear. Also, as
shown in \cite{SmithArdakani2010}, under mild conditions, $I$, as
approximated in (\ref{eqn:itapprox}), is a convex function of $\Lambda$. 

Given an SNR, the inner-code optimization is performed over three loops,
iterating over discrete sets of values for each of $d_c$, $\nu$, and $L_0$
(see Sec.~\ref{subsubsec:discretization}). Once the values of $d_c$, $\nu$,
and $L_0$ are fixed, the problem of designing complexity-optimized
inner-code becomes convex, and can be solved by the method described in
Sec.~\ref{subsubsec:optimizationalgorithm}.

Once all three loops are executed, the ensemble with lowest complexity,
according to (\ref{eqn:icomp}), is chosen as the inner-code ensemble. The
loop over the outer-code table then outputs the staircase-code
inner-code-ensemble pair that achieves the minimum  overall complexity,
according to (\ref{eqn:eta}), as the optimized concatenated code.

\subsection{Practical Considerations}
\label{subsec:practicalconsiderations}

\subsubsection{Discretization}
\label{subsubsec:discretization}

In practice, the  integral in (\ref{eqn:itapprox}) is estimated by a sum
over a quantized version of the $[p_t,p_0]$ interval. Let $Q$ be the
number of quantization points. Define $\Delta \triangleq (p_0-p_t)/Q$ and
let
\[
q_i = p_t + i \Delta, \quad i \in \{0 , 1, \ldots , Q-1\}.
\]
We define a discrete approximation of the integral in (\ref{eqn:itapprox}) as
\[
I_Q = \sum_{i=0}^{Q-1} \frac{\Delta}{q_i \ln ( \frac{q_i}{ f_{\Lambda}(q_i) })},
\]
which we use in the objective function in (\ref{eqn:opt}), instead of the
integral. The constraint $f_{\Lambda}(q_i) < q_i \quad i \in \{0 , 1,
\ldots, Q\}$ then ensures the openness of the EXIT-curve throughout the
decoding procedure.

Similarly, intervals $[0,\nu^{\text{max}}]$ and $[0,L_0^{\text{max}}]$, are
quantized with $Q_{\nu}$ and  $Q_{L_0}$ points when searching over values
of $\nu$ and $L_0$  at the inner-code ensemble optimization. Here,
$L_0^{\text{max}}$ was defined in (\ref{eqn:L0max}) and
$\nu^{\text{max}}$ 
 is the maximum value for $\nu$
that we consider in our search  for the optimal inner-code
ensemble. The values of $Q$,
$Q_{L_0}$ and $Q_{\nu}$ allow the designer to trade-off between accuracy
and computational complexity of the design process.

\subsubsection{Optimization Algorithm}
\label{subsubsec:optimizationalgorithm}

Even when $d_c$, $\nu$, and $L_0$ are fixed, the objective function is
non-linear and is not easily differentiable. To solve the optimization
problem, we used the sequential quadratic programming (SQP) method
\cite{powell1978fast}. This method is an iterative procedure, at each
iteration of which a quadratic model of the objective function is optimized
(subject to the constraints), and the solution is used to construct a new
quadratic model of the objective function. During this procedure, we update
$p_t$ whenever $\Lambda$ undergoes a large change (see
(\ref{eqn:itapprox})--(\ref{eqn:L0}) and \cite[Sec.~II.B]{SmithArdakani2010}).

An issue with using the SQP algorithm is that it needs to be initialized
with a feasible point. In our design procedure, we initialize the algorithm
with the parameters of the rate-optimized ensemble
\cite{SmithArdakani2010}.

\subsubsection{Interleaving Between Inner and Outer Code}
The outer staircase
code threshold $p_{\text{sc}}$ is computed assuming that the
outer code sees a binary symmetric channel, i.e., a channel
with independent and identically distributed bit errors occurring with
probability $p_{\text{sc}}$.  The inner decoder, however, produces
correlated errors.
To mitigate the error correlation, we
use a \textit{diagonal interleaver}
as in \cite{ZhangKschischang2017}. 
We suppose that each staircase block is of size $M^2$,
and we choose the inner code dimension $k_{\text{in}}$ to
divide $M^2$.
We define the \emph{packing ratio}, $m$, as the number of inner codewords
associated with a staircase block, i.e., $m = M^2/k_{\text{in}}$.

\begin{table}[t!]
\caption{Quantifying Finite Interleaving Loss}
\label{tbl:VarR}
\centering
\begin{tabular}{r|cccc}
Packing Ratio $m$     & 1 & 2 & 4 & $\geq 8$ \\\hline
Performance Loss (dB) & 0.02 & 0.01 & 0.007 & $\approx 0$
\end{tabular}
\end{table}

Table~\ref{tbl:VarR} shows the performance loss, relative
to ideal interleaving, obtained
for different packing ratios, assuming an outer staircase
code of rate 15/16 with $M=704$ and using an inner code sampled
from an optimized ensemble.  The ideal interleaving
threshold was estimated by interleaving inner codewords over
multiple staircase blocks.  At packing ratios exceeding 4,
the performance degradation becomes negligible, justifying
the use of the simple binary symmetric channel BER analysis of staircase codes.

\begin{figure}[t!]
\centering
\includegraphics{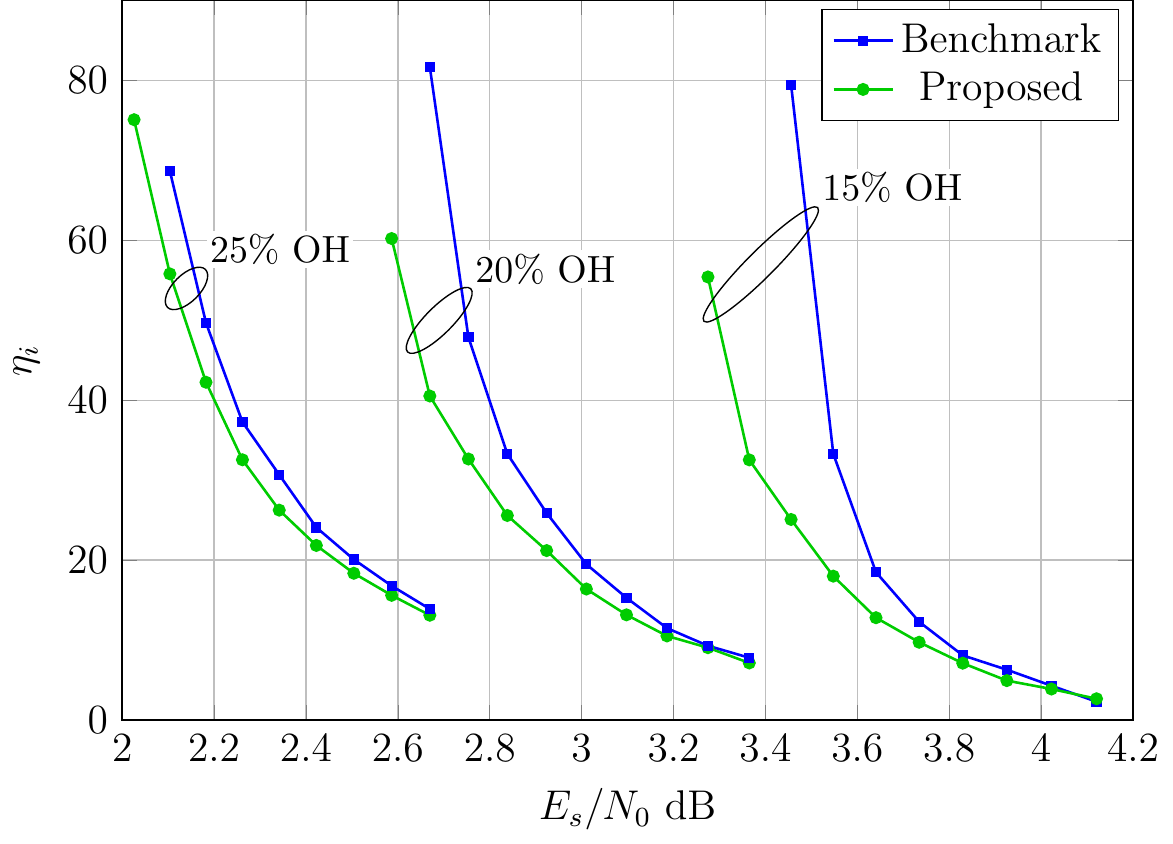}
\caption{The $(E_s/N_0,\eta_i)$ Pareto frontiers of the inner code in the
proposed design, compared with the benchmark design of
\cite{ZhangKschischang2017}, at 15\%, 20\%, and 25\% OHs.}
\label{fig:pf}
\end{figure}

\section{Results}
\label{sec:results}

\subsection{Pareto Frontier}
\label{subsec:paretofrontier}
We searched staircase codes of \cite[Table~1]{ZhangKschischang2017} for the
optimal outer code. We refer the reader to \cite{zhang2014staircase} to
see how these codes are obtained. The reader should note that there is a
slight difference between two of the entries in the earlier table
\cite[Table~1]{zhang2014staircase} (which included $t=5$-error-correcting
component codes)
and the later table
\cite[Table~1]{ZhangKschischang2017} (which includes only results
corresponding to the more practical $t=4$ component codes).

We used the following parameters in designing inner-code ensembles: $D_v =20$, $
\nu^{\text{max}}=4$, $Q=200$, $Q_{\nu}=40$. We chose $Q_{L_0}$ such that
step-sizes of the quantized version of $[0,L_0^{\text{max}}]$ are no
greater than 0.01. We used the sum-product algorithm in generating the
elementary EXIT charts, and $10^6$ samples were produced at each pass of
the Monte-Carlo simulation.

Fig.~\ref{fig:pf} shows the $(E_s/N_0,\eta_i)$ Pareto frontier for the
designed inner-codes, at 15\%, 20\%, and 25\% OHs. The  Pareto frontiers
are also compared with those of \cite{ZhangKschischang2017}. Similar to
\cite{ZhangKschischang2017}, all our concatenated code designs picked the
highest-rate staircase code available, with $R_{\text{sc}} = 15/16$ and
$p_{\text{sc}} = 5.02 \times 10^{-3}$. As can be seen from
Fig.~\ref{fig:pf}, the proposed design outperforms the design in
\cite{ZhangKschischang2017}. The inner codes of this paper achieve the
performance of the inner codes of \cite{ZhangKschischang2017}, with up to
71\%, 50\%, and 19\% reduction in complexity, at 15\%, 20\%, and  25\% OHs,
respectively. Also, compared to \cite{ZhangKschischang2017}, the designed
concatenated codes operate at up to 0.23 dB, 0.14 dB, and 0.06 dB closer to
the constrained Shannon limit, at 15\%, 20\%, and  25\% OHs, respectively.

To study the performance of the designed inner codes at overall 20\% OH, we
sampled codes of length up to 100,000 from each of the complexity-optimized
inner-code ensembles. We simulated the transmission of codewords over an
AWGN channel. Codewords were decoded using the sum-product algorithm with
floating-point message-passing, and the code performance was obtained by
averaging the codeword BERs. Note that we only care about the BER of the
information set of a codeword.

\begin{figure}[t!]
\centering
\includegraphics{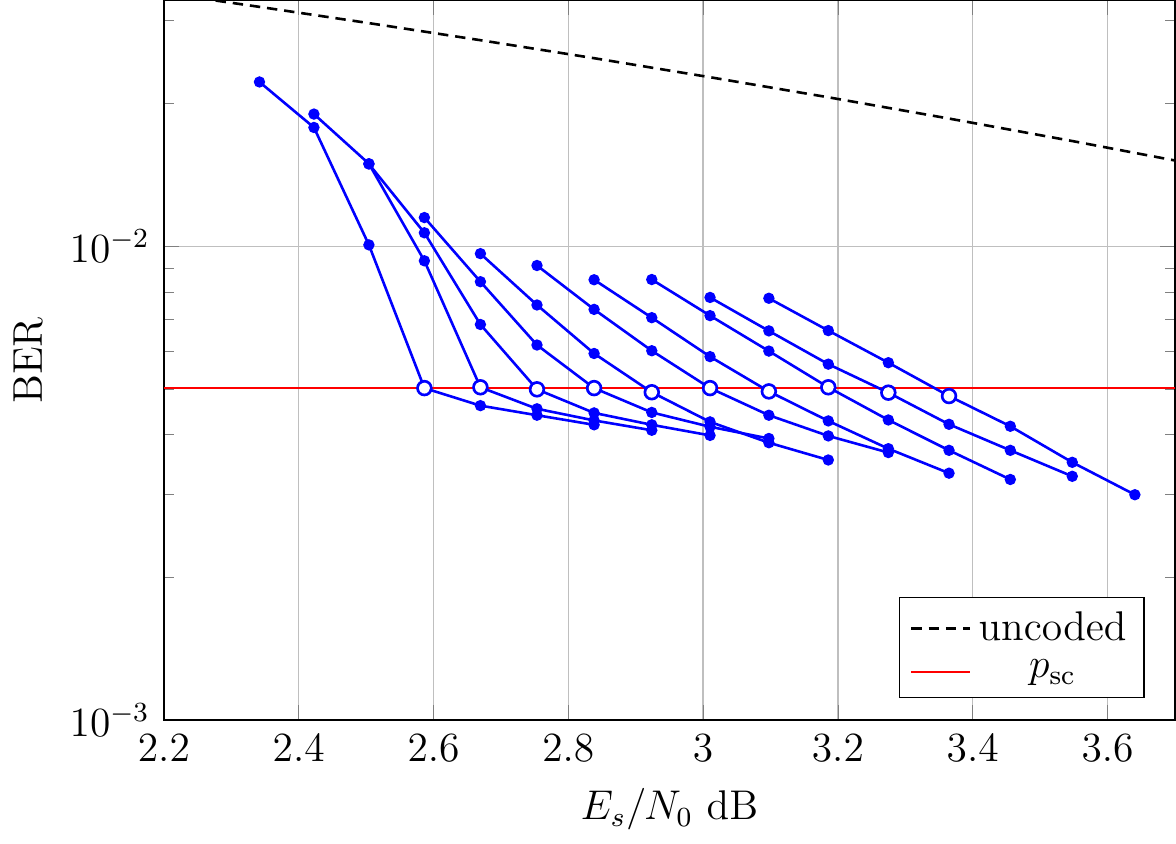}
\caption{Simulated decoder BERs of inner codes, sampled from the
complexity-optimized ensembles, for designs at 20\% OH. The mid-point on each BER curve 
(highlighted by an `o') is the code operational point, i.e, the SNR for 
which the inner code is designed to achieve $\text{BER} \leq p_{\text{sc}}$.} 
\label{fig:simber}
\end{figure}

In Fig.~\ref{fig:simber}, obtained BERs are plotted versus SNR. The
$p_{\text{sc}}$ line shows the outer staircase code threshold. The
mid-point SNR on each curve (highlighted by on `o') is  the code operational point, i.e., the SNR
for which the code is designed. Note that all BERs of the sampled codes hit
at, or below, the outer-code threshold, at their operational point,
verifying our design approach.

\subsection{Two Design Examples}
\label{subsec:twodesignexamples}
Here we present two interesting examples of the complexity-optimized
concatenated code designs at 20\% OH. In both of these examples, the outer
code picked was the $R_{\text{sc}} = 15/16$ and $p_{\text{sc}} = 5.02
\times 10^{-3}$ staircase code. 

Example 1: An FEC scheme operating at 1.27 dB from the
constrained Shannon limit. The optimization procedure yields the following
ensemble for the inner code:
\begin{gather*}
L(x) = 0.1556+0.1389x+0.2941x^3+0.4113x^4,\\
R(x) = x^{24}.
\end{gather*}
This code requires a maximum of 9 iterations to bring the BER below the outer-code threshold, which gives an inner-code complexity score of 25.59.

Example 2: An FEC scheme operating at 1 dB from the constrained
Shannon limit. The optimization procedure yields the following ensemble for
the inner code:
\begin{gather*}\
L(x) = 0.1480+0.1111x+0.4539x^3+0.0911x^4+0.0973x^6+0.0985x^7,\\
R(x) = x^{28}.
\end{gather*}
This code requires a maximum of 18 iterations to bring the BER below the outer-code threshold, which gives an inner-code complexity score of 60.24.

\subsection{Comparison to Other Works}
\label{subsec:comparisontootherworks}
\begin{figure}[t!]
\centering
\includegraphics{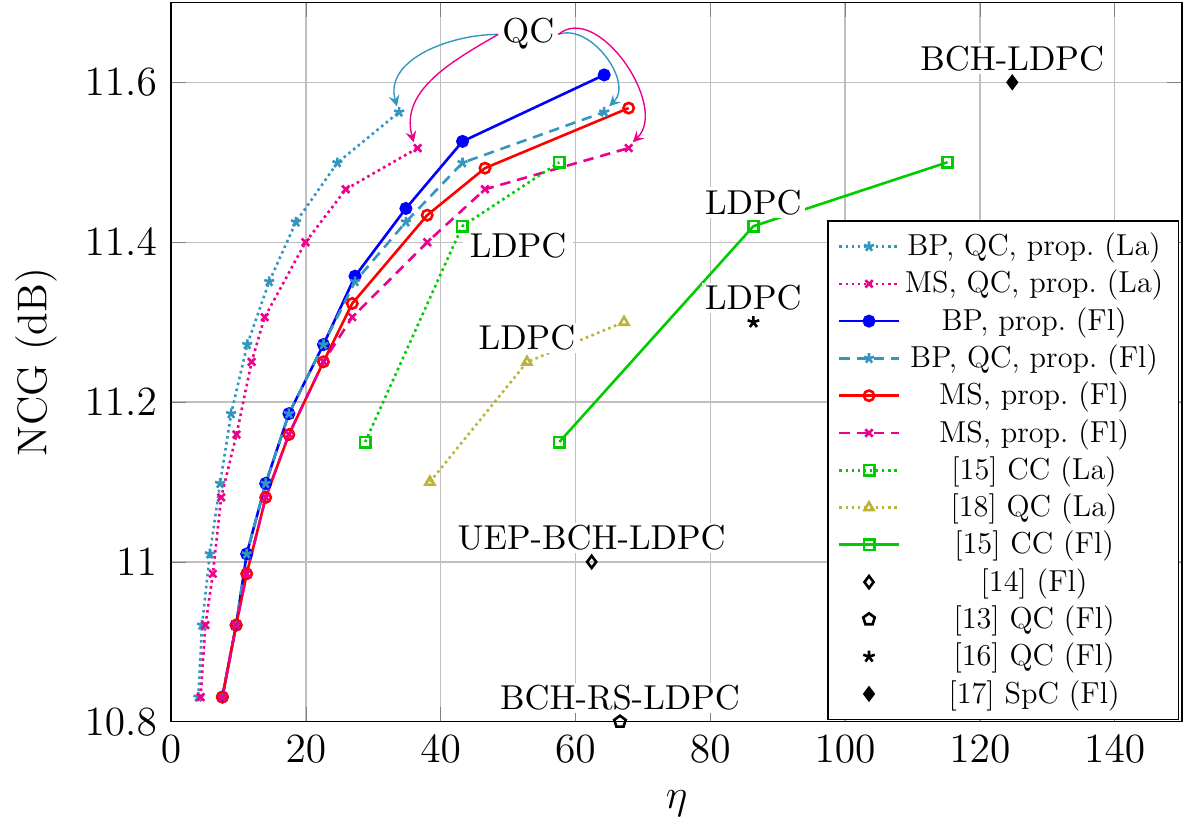}
\caption{NCG and $\eta$ comparisons of the proposed concatenated design and
other soft-decision FEC schemes, at 20\% OH.  Decoders using a
flooding (resp., layered) decoding schedule are denoted with
Fl (resp. La).
For the proposed codes
(denoted as ``prop.''), the inner decoding algorithm (MS or BP)
is specified. Block length 30000 is
considered for the designs with quasi-cyclic (QC) structured inner codes. The
following abbreviations are used in describing the referenced codes. BCH:
Bose---Ray-Chaudhuri---Hocquenghem, UEP: Unequal Error Protection, RS: Reed-Solomon,
CC: Convolutional Code, SpC: Spatially Coupled.} \label{fig:compete}
\end{figure}

To compare our work with the existing designs, in Fig.~\ref{fig:compete}, we
have plotted the NCG versus complexity, at 20\% OH, for our designed codes, and
also for several other existing FEC solutions. Since the referenced code
designs are based on min-sum (MS) or offset-MS decoding, we also simulated the
obtained inner codes using the offset-MS algorithm with unconditional
correction \cite{zhao2005implementation}.

Compared to code designs decoded under a flooding schedule, the obtained
MS-based codes achieve, at similar complexities, a 0.77 dB gain over the
code in \cite{SugiharaOnoharaLDPC2010}, a 0.57 dB gain over the code in
\cite{Mizuochi2013}, and a 0.42 dB gain over the code in
\cite{ChangConvLDPC2012}. The designed codes achieve NCGs of codes in
\cite{ChangConvLDPC2012} and \cite{MoreroLDPC2011} with more than a 56\%
reduction in complexity, and the excellent NCG of code in
\cite{sugihara2016scalable} with 46\% reduction in complexity. 

Compared to code designs where the inner code is decoded under a layered
schedule, the obtained MS-based codes achieve NCGs of codes in
\cite{ChangQCLDPC2011} with more than 57\% reduction in complexity, and
achieves NCGs of codes in \cite{ChangConvLDPC2012} with 15\% to 41\%
reduction in complexity.

While some designs in \cite{ChangConvLDPC2012}, decoded under a layered
schedule, come close to the proposed MS-based codes, the proposed BP-based
codes, decoded under flooding schedule, strictly dominate the existing
designs. The BP-based codes achieve the NCGs of the existing designs with
at least 62\% and 24\% reduction in complexity compared to code designs
decoded under a flooding schedule and layered schedule, respectively. The
BP-based codes achieve at least 0.45 dB and 0.11 dB greater NCG over the
existing designs, at nearly the same $\eta$, compared to code designs
decoded under flooding schedule and layered schedule, respectively. 

The latency of the proposed concatenated code can be obtained by adding the
latencies of the inner and the outer codes.  The latency is dominated
by the staircase decoder.  For example, at $200$~Gb/s,
for a staircase block containing $4.65 \times 10^5$ information bits
and a staircase decoding window size $W=6$,
the decoding latency of the proposed concatenated code
(including the inner code)
is $\approx 2.8 \times 10^6$ bit periods, or 14~$\mu$s, which
is an acceptable latency in many OTN applications.

\subsection{Quasi-Cyclic-Structured Inner Codes}
\label{subsec:quasicyclicstructuredinnercodes}

The inner codes considered so far have been randomly structured and have
large block lengths. Decoder architectures for such codes are often plagued
with routing and message-permutation complexities. In order to obtain a
more pragmatic implementation of the proposed FEC scheme, we adopt a
quasi-cyclic (QC) structure for the inner codes.  The QC structure is well
known to be hardware-friendly and leading to energy-efficient
implementations;  see \cite[Ch.~3]{Milicevic2017Low} and references
therein.

We constructed girth-8 inner-codes of coded component length $30000
\pm$1\%, based on the obtained inner-code ensembles, for the concatenated
code at 20\% OH. As can be seen from Fig.~\ref{fig:compete}, the
concatenated FEC performs as well, with QC-structured inner-codes, as with
randomly structured inner-codes, with only a small loss in performance when
operating at a high NCG. Note that, however, we do not make any claim of
optimality for the code constructions with QC-structured inner-codes, as
the optimization procedure used assumes a random structure for the
inner-code coded component.

The structure of the QC codes also allows for layered decoding of the
constructed inner codes. As can be seen from Fig.~\ref{fig:compete}, the
concatenated scheme with inner-code coded component length $30000 \pm$1\%,
decoded under layering schedule, performs at up to 50\% lower complexity
compared to the scheme with the inner-code coded component, decoded under
flooding schedule. Compared to the existing code designs decoded
under a layered schedule, the designed codes, with QC inner-codes decoded
under layering schedule, achieve a similar NCG with at least 40\%
reduction in complexity.

\begin{figure}[t!]
\centering
\includegraphics{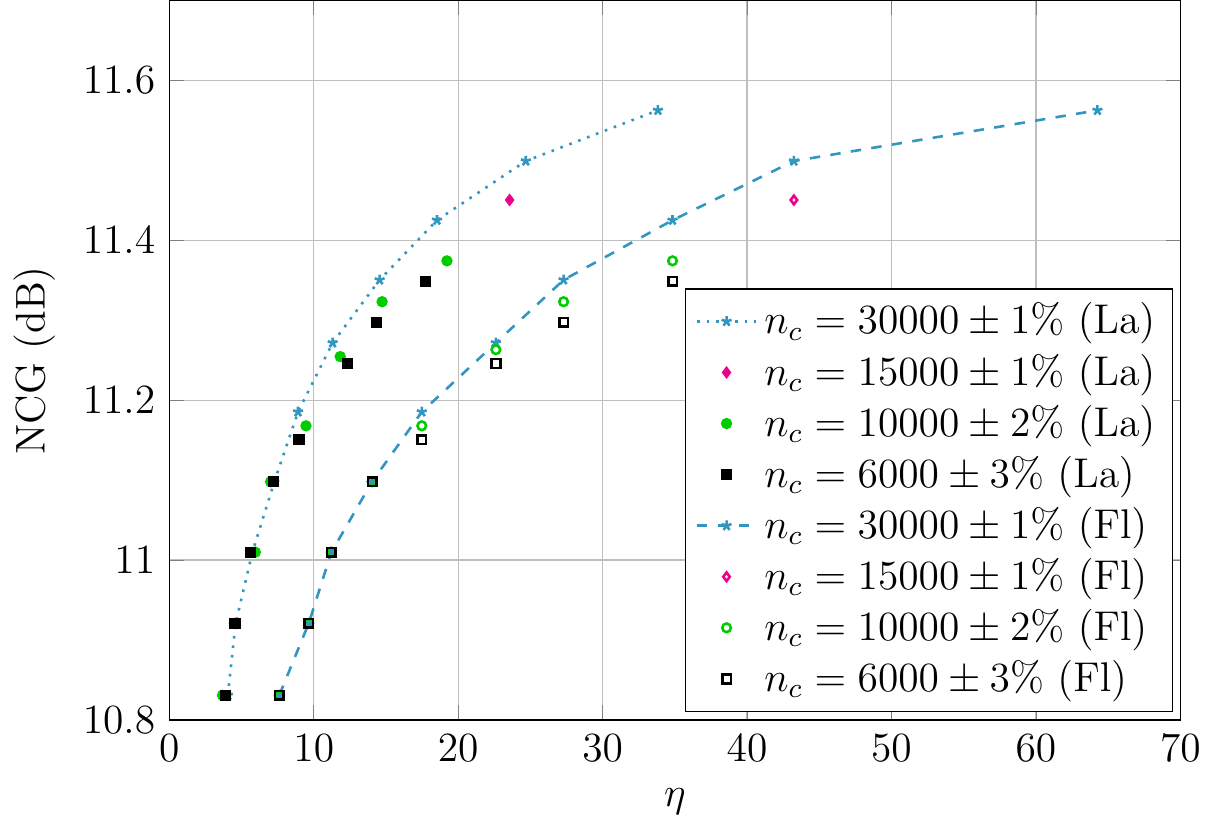}
\caption{NCG and $\eta$ comparisons of the QC constructions of the designed
concatenated FEC, at 20\% OH, under layered (La) and
flooding (Fl) schedules.}
\label{fig:innercomp}
\end{figure}

While a length $30000$ LDPC code can be considered practical for OTN
applications \cite{ChangConvLDPC2012}, we have also constructed
QC-structured inner-codes of shorter lengths ($6000\pm3\%$, $10000\pm2\%$,
and $15000\pm1\%$) and possibly lower girths, based on the obtained
inner-code ensembles, at 20\% OH. Note that, according to (\ref{eqn:eta}) and
(\ref{eqn:icomp}), using a short inner code does not change the complexity
score of the overall code; however, having a short inner code leads to a more
practical implementation, as it greatly reduces wiring and routing
complexities.  A comparison between the concatenated FEC scheme with inner-code
coded component of various lengths is provided in Fig.~\ref{fig:innercomp}. 

As can be seen from Fig.~\ref{fig:innercomp}, when shorter inner codes are used, the loss in NCG is not
significant, although the loss becomes bigger, as the NCG increases or as the inner-code length becomes shorter.
Nevertheless, schemes with inner-code coded component of length
$6000\pm3\%$, decoded under a layered schedule, operate at up to 50\%
less complexity, compared to schemes with an inner-code coded component of
length $30000\pm1\%$, decoded under a flooding schedule.

\section{Conclusion}
\label{sec:conclusion}

In this paper we have proposed a concatenated code design that improves
significantly upon the results of \cite{ZhangKschischang2017}. The
complexity-optimized error-reducing inner code, concatenated with an outer
staircase code, forms a low-complexity FEC scheme suitable for high
bit-rate optical communication. An interesting feature that emerges from
the inner-code optimization is that a fraction symbols are better left
uncoded, and only protected by the outer code. We showed that, compared to
\cite{ZhangKschischang2017}, with this modified design, the inner-code
complexity can be reduced by up to 70\%. We showed that the concatenated
code designs have lower complexity than, to the best of our knowledge, any
other existing soft-decision FEC scheme. 

To realize a pragmatic and energy-efficient implementation for the proposed
FEC scheme, we constructed QC inner codes, based on the obtained ensembles.
We showed that, QC-structured inner codes  with practical lengths can
achieve the performance of the randomly constructed inner codes. We
simulated layered decoding of the QC inner codes and showed that with
layered decoding, the complexity score of the FEC scheme can be reduced by
up to 50\%.

There are two worthwhile directions for further research. First, using an
outer staircase code with higher rate is likely to yield concatenated code
designs with even lower complexity. To test this hypothesis, we
extrapolated the table of staircase codes in \cite[Table
1]{ZhangKschischang2017}, and observed that at 20\% overall OH, $\eta$
continues to decrease as the outer staircase code OH decreases down to 2\%.
Unfortunately, it is not trivial to design and implement staircase codes
with a very high rate, because the staircase block size becomes very large
as the code rate increases.

Second, as shown in this work, using a layered schedule in decoding
significantly reduces the decoding complexity. However, we cannot make any
claim of optimality for the proposed codes when decoded under a layered
schedule. A topic of future work is to modify the inner-code design
procedure to obtain complexity-optimized inner codes that are decoded under
a layered schedule.

\end{document}